\documentclass[lettersize,journal]{IEEEtran}
\usepackage{amsmath,amsfonts}
\usepackage{algorithm}
\usepackage{array}
\usepackage{bm}
\usepackage{booktabs}
\usepackage{textcomp}

\usepackage{subcaption}
\usepackage{makecell} 
\usepackage{subfloat}
\usepackage{caption}
\usepackage{booktabs}
\usepackage{multirow}
\usepackage{algorithmic}
\usepackage{amssymb}
\usepackage{booktabs}
\usepackage{graphicx}
\usepackage{stfloats}
\usepackage{url}
\usepackage{hyperref}
\usepackage{verbatim}
\usepackage{cite}
\begin{document}

\title{Uncertainty-Driven Hierarchical Sampling for Unbalanced Continual Malware Detection with Time-Series Update-Based Retrieval}

\author{Yi Xie, Ziyuan Yang, Yongqiang Huang, Yinyu Chen, Lei Zhang, Liang Liu, Yi Zhang, ~\IEEEmembership{Senior Member,~IEEE,}
\thanks{This work was supported in part by the National Natural Science Foundation of China under Grant 62271335; and in part by the Sichuan Science and Technology Program under Grant 2025ZNSFSC0470. \textit{(Corresponding author: Yi Zhang)}}
\thanks{Yi Xie, Yongqiang Huang, Lei Zhang, Liang Liu, and Yi Zhang are with the School of Cyber Science and Engineering, Sichuan University, Chengdu 610065, China (e-mail: 
yix@stu.scu.edu.cn; yqhuang2912@gmail.com; zhanglei2018@scu.edu.cn; liangzhai118@scu.edu.cn; yzhang@scu.edu.cn)}
\thanks{Ziyuan Yang and Yinyu Chen are with the College of Computer Science, Sichuan University, Chengdu 610065, China (e-mail: cziyuanyang@gmail.com; cyy262511@gmail.com).}
}

\markboth{Journal of \LaTeX\ Class Files,~Vol.~14, No.~8, August~2021}%
{Shell \MakeLowercase{\textit{et al.}}: A Sample Article Using IEEEtran.cls for IEEE Journals}


\maketitle
\begin{abstract}

Android malware detection continues to face persistent challenges stemming from long-term concept drift and class imbalance, as evolving malicious behaviors and shifting usage patterns dynamically reshape feature distributions. Although continual learning (CL) mitigates drift, existing replay-based methods suffer from inherent bias. Specifically, their reliance on classifier uncertainty for sample selection disproportionately prioritizes the dominant benign class, causing overfitting and reduced generalization to evolving malware.
To address these limitations, we propose a novel uncertainty-guided CL framework. First, we introduce a hierarchical balanced sampler that employs a dual-phase uncertainty strategy to dynamically balance benign and malicious samples while simultaneously selecting high-information, high-uncertainty instances within each class. This mechanism ensures class equilibrium across both replay and incremental data, thereby enhancing adaptability to emerging threats.
Second, we augment the framework with a vector retrieval mechanism that exploits historical malware embeddings to identify evolved variants via similarity-based retrieval, thereby complementing classifier updates.
Extensive experiments demonstrate that our framework significantly outperforms state-of-the-art methods under strict low-label conditions (50 labels per phase). It achieves a true positive rate (TPR) of 92.95\% and a mean accuracy (mACC) of 94.26\%, which validates its efficacy for sustainable Android malware detection.

\end{abstract}


\section{Introduction}
\IEEEPARstart{T}{he} rapid and continuous evolution of Android malware presents a critical challenge to modern cybersecurity defenses. While machine learning-based Android malware detection models have achieved notable success in static evaluations, their performance often degrades over time due to concept drift.
This degradation arises as both malicious and benign software behaviors evolve dynamically. Malware developers employ techniques such as polymorphic obfuscation, dynamic payload injection, and API call pattern mutation to evade detection, whereas benign applications continually introduce new features through regular updates, thereby altering their observable characteristics. Addressing these evolving distributions necessitates malware detectors that can continually adapt in order to sustain robust performance over time.

Continual learning (CL) has emerged as a promising solution for sustainable Android malware detection, which enables models to be incrementally updated as new data becomes available. Previous works, such as that by Chen \textit{et al.}~\cite{chen2023continuous}, have investigated the integration of contrastive learning with active learning for sample selection, which aims to preserve detection accuracy during continual updates. Nevertheless, these methods encounter inherent limitations, primarily because they rely on classifier uncertainty or pseudo-loss scores as the principal criteria for sample selection during incremental updates.

One of the major challenges inadequately addressed by existing methods is the severe class imbalance in real-world Android malware detection ~\cite{kim2023automated,wang2025archsentry}, where benign samples vastly outnumber their malicious counterparts. Uncertainty-driven sample selection methods inherently bias toward the benign majority, as these samples dominate the feature space and influence uncertainty estimates. This bias leads to suboptimal replay sets, overfitting to benign classes, and degraded capacity to generalize against emerging malware behaviors.

To overcome these limitations, we propose a novel Uncertainty-Guided Sampling and Retrieval~(UGSR) framework for continual Android malware detection. Specifically, UGSR introduces two core innovations: one to balance class distributions during the sampling stage and another to enhance adaptability to evolving malware during the detection stage.

For the sampling stage, we propose a hierarchical uncertainty sampling strategy that decouples the sampling process from the core detection model.
Unlike conventional binary sampling methods that only distinguish between malicious and benign instances, our hierarchical sampler additionally operates on fine-grained malware family classes alongside benign applications. By independently estimating uncertainty at both the multi-class and binary levels, our hierarchical sampler ensures the selection of high-information and high-uncertainty samples, while simultaneously maintaining class balance. Furthermore, the hierarchical design enhances intra-class diversity, especially among malware families, thereby ensuring a more representative and informative sample set for continual updates.

At the detection stage, we introduce an adaptive vector retrieval mechanism that reformulates the detection task as a similarity-based retrieval problem. By constructing and continuously updating a class-balanced codebook that contains only the top 50 representative benign and malware embeddings, the system can detect evolved malware variants via vector similarity rather than solely relying on classifier predictions. To further improve retrieval robustness, we incorporate the Equiangular Tight Frame (ETF) principle to maximize inter-class separation and promote intra-class compactness within the codebook, reducing false positives and enhancing detection accuracy.

In summary, the main contributions of this work are as follows:
\begin{itemize}
    \item We propose UGSR, a novel continual learning framework that decouples uncertainty-guided sample selection from Android malware detection, mitigating the limitations of shared-network approaches and improving detection robustness.
    \item We introduce a hierarchical uncertainty sampling strategy that dynamically balances benign and malware classes while enhancing intra-class diversity, effectively addressing the class imbalance problem in continual Android malware detection.
    \item We develop an adaptive vector retrieval mechanism with evolutionary codebook update and ETF-based feature regularization, which enables effective detection of malware variants while reducing false positives.
\end{itemize}

\section{Related Works}
\subsection{Uncertainty sampling}
Uncertainty sampling is a foundational technique in active learning, aimed at selecting the most informative samples by quantifying model uncertainty in its predictions. Traditional methods such as entropy-based sampling~\cite{wiewel2021entropy}, least confidence~\cite{settles2009active}, and margin sampling~\cite{gal2016uncertainty} have been widely adopted in image classification and text processing tasks. For instance, Settles \textit{et al.}~\cite{settles2009active} demonstrated that margin-based selection accelerates model convergence by focusing on samples near decision boundaries. 

With the advent of deep learning, uncertainty estimation methods have evolved. Gal \textit{et al.}~\cite{gal2016uncertainty} introduced Monte Carlo  Dropout to approximate Bayesian uncertainty in neural networks, which enables efficient sample selection in image segmentation. Similarly, Beluch \textit{et al.}~\cite{beluch2018power} showed that ensemble-based methods outperform single-model approaches in medical image analysis, highlighting the importance of model diversity.

Recent studies have explored hybrid uncertainty sampling strategies. Gissin \textit{et al.}~\cite{gissin2019discriminative} combined discriminative models with uncertainty-driven sampling to reduce annotation costs in object detection tasks. Do \textit{et al.}~\cite{do2021semi} leveraged adversarial perturbations to identify uncertain regions in latent spaces for semi-supervised learning. Extending these ideas to continual learning, Ahn \textit{et al.}~\cite{ahn2019uncertainty} proposed incorporating temporal consistency metrics to adapt uncertainty sampling to dynamic data streams. This approach addresses concept drift and provides particularly valuable insights to evolve Android malware detection, as subsequently validated in Barbero \textit{et al.}~\cite{barbero2022transcending}. 

However, most existing uncertainty sampling methods focus on static domains, such as image or text classification. Their direct application to sequential, non-visual tasks like Android malware detection remains underexplored, especially when considering challenges such as extreme class imbalance and the need for long-term adaptability.

\subsection{Continual learning}
Continual learning (CL), also known as incremental learning, addresses the challenge of updating models on non-stationary data streams while mitigating catastrophic forgetting. Classic CL approaches include regularization-based methods, such as EWC~\cite{kirkpatrick2017overcoming}, which preserves important parameters from previous tasks, and architectural strategies like Progressive Neural Networks~\cite{rusu2016progressive}, which expand model capacity incrementally. Rehearsal-based techniques, such as iCaRL~\cite{rebuffi2017icarl}, leverage exemplar storage to preserve information from past classes. 

Recent CL methods predominantly focus on balancing model adaptability and memory efficiency. DER~\cite{yan2021der} dynamically expands feature representations for new tasks while freezing older parameters, whereas PODNet~\cite{douillard2020podnet} employs distillation loss to preserve knowledge across incremental phases.

In the context of Android malware detection, continual learning is critical due to the rapid evolution of threats. Pang \textit{et al.}~\cite{pang2021toward} introduced a rehearsal-based CL framework for malware classification that utilizes a small memory buffer to preserve representative samples from older families. Shin \textit{et al.}~\cite{shin2024towards} proposed a meta-incremental approach that trains models to generalize better to unseen malware classes while addressing inherent class imbalance. Gong \textit{et al.}~\cite{gong2024improving} integrated adversarial training with knowledge distillation to enhance model robustness against evolving features. 

Despite these advances, most CL methods for malware detection focus primarily on class-incremental scenarios and often overlook the interplay between feature drift, class imbalance, and temporal dependency in malware behaviors. This gap motivates the development of our framework’s dual-level uncertainty-guided sampling and retrieval mechanism, which explicitly addresses these challenges.

\subsection{Continual few-shot learning for malware detection}
Few-shot class-incremental learning (FSCIL) has garnered increasing attention as a promising paradigm for real-world malware detection, where labeled data is scarce and new threat variants continuously emerge~\cite{tao2020fewshot,ren2019incremental}. While initially introduced by Tao \textit{et al.}~\cite{tao2020fewshot} for general incremental learning scenarios, FSCIL faces unique challenges when adapted to Android malware detection, due to the dynamic nature of cyber threats and the severe scarcity of labeled attack samples.

Several FSCIL methods have been proposed to mitigate catastrophic forgetting and feature drift. TOPIC~\cite{tao2020fewshot} leverages a growing neural gas network to model feature topology, which preserves learned representations while incorporating new threat classes. FSLL~\cite{mazumder2021few} reduces computational overhead by selectively updating only a subset of model parameters during incremental phases, which is crucial for resource-constrained Android malware detection systems. By enhancing feature scalability, SPRR~\cite{zhu2021self} effectively supports incremental learning systems - a critical capability given malware authors' rapidly adapting attack strategies. Additionally, Cheraghian \textit{et al.}~\cite{cheraghian2021semantic} introduced semantically informed embeddings to improve knowledge retention for known malware families, which mitigates forgetting in long-term learning scenarios.

Despite these advances, applying FSCIL to Android malware detection remains challenging. Frameworks such as IMC have demonstrated the feasibility of classifying unknown malware classes while maintaining accuracy on known threats. However, IMC and similar approaches are limited to single-session incremental learning and fail to fully address the continuous, long-term evolution of malware families. Unlike applications in image recognition~\cite{qiang2022incremental,zhou2022forward,dong2021few,achituve2021gptree} or vehicle recognition~\cite{li2022few}, Android malware detection faces unique challenges, such as decision boundary confusion caused by polymorphic malware variants and catastrophic forgetting of older malware families over extended learning cycles. 

Most existing FSCIL methods show promising results in controlled settings but struggle to maintain stability and accuracy under real-world conditions, where new malware variants emerge frequently and labeled samples are scarce. This persistent gap highlights the need for frameworks such as ours, which explicitly balance class distributions, enhance intra-class diversity, and dynamically adapt through uncertainty-guided sampling and vector retrieval mechanisms.

\section{Methodology}   
\subsection{Overview}

\begin{figure*}[!t]
\includegraphics[width=\textwidth]{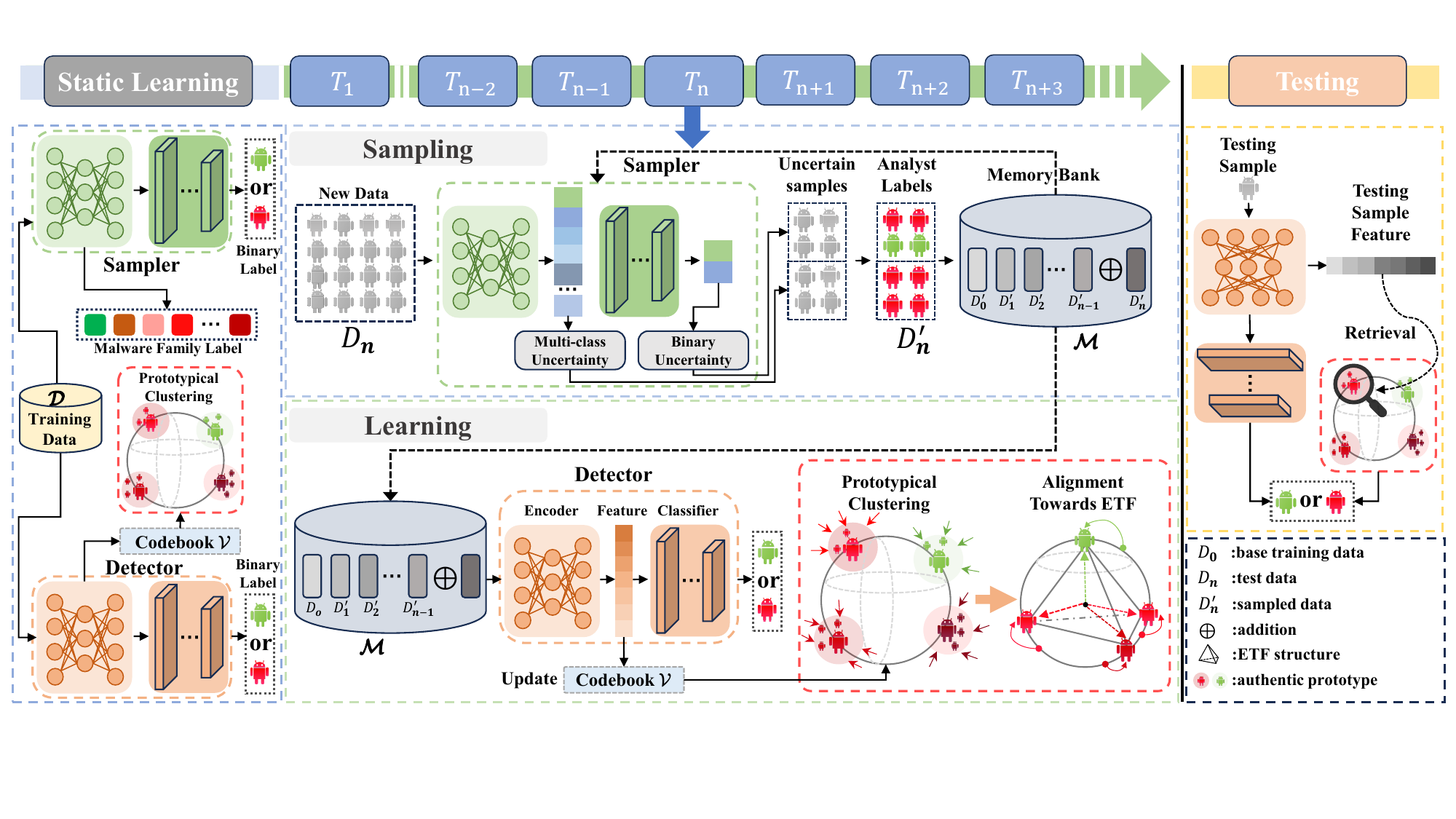}
\caption{The overall pipeline of our proposed method.}
\label{fig1}
\end{figure*}

In this work, we propose a novel sustainable Android malware detection method, UGSR, for continual learning scenarios. An overview of the proposed UGSR framework is illustrated in Fig.~\ref{fig1}. The training process of UGSR consists of three stages: static learning, continual learning, and testing. The continual learning stage includes two phases: sampling and learning. Initially, we perform static training using the available data to obtain an initial model. At each continual learning stage, we first use a hierarchical uncertainty sampler to select representative samples for subsequent stages. In the learning phase, the primary challenge is to preserve the knowledge acquired earlier while incorporating new knowledge from novel samples. In the testing stage, we reformulate the detection task as a retrieval problem to prevent overfitting to the dominant class.

\subsection{Static Learning}
Given the input data and corresponding labels $(x; y^{\text{mul}}; y^{\text{bin}})$, where each input data $x$ is associated with two labels: $y^{\text{mul}} \in \{0,1\}^{C+1}$ (including one benign and $C$ fine-grained malware family classes), and $y^{\text{bin}} \in {\{0,1\}}$, which represent one-hot encoded multi-class labels and a binary Android malware detection label for $x$, respectively. 
The training process involves sequentially learning two components: the sampler $\mathcal{F}_{\text{S}}$ and the detector $\mathcal{F}_{\text{D}}$. 

\noindent \textbf{Sampler $\mathcal{F}_{\text{S}}$:} 
The sampler $\mathcal{F}_{\text{S}}$ is a two-stage architecture consisting of a multi-class module $\mathcal{F}_{\text{mul}}$, which provides reliable fine-grained multi-class Android malware detection, and a binary-class module $\mathcal{F}_{\text{bin}}$, which produces the final binary detection score. The two-stage operation can be formulated as:

\begin{equation}
    f_m = \mathcal{F}_{\text{mul}}(x).
\end{equation}
\begin{equation}
     \mathcal{F}_{\text{S}}(x) = \mathcal{F}_{\text{bin}}(f_m).
\end{equation}
Specifically, for the multi-class module $\mathcal{F}_{\text{mul}}$, we utilize Dempster-Shafer Theory of Evidence \cite{sensoy2018evidential, chen2024evidence, chen2023evil, yang2025patient} to generate reliable fine-grained multi-class malware predictions. The logits $f_m$ can be regarded as the parameters of a Dirichlet distribution, which models the density of malware probability. The density function is defined as follows:
\begin{equation}
    D(p_m \mid f_m) =  \Bigg\{
    \begin{array}{ll}
        \frac{1}{B(f_m)} \prod_{c=1}^C p_m^{f_m^c-1} & \text{for } p_m \in \mathcal{S}^{C}, \\
        0 & \text{otherwise},
    \end{array}
    \Bigg.
\end{equation}
where $p_{m}$ represents the multi-class malware probability, $B(f_m)$ is the $C$-dimensional multinomial beta function for the parameter $f_m$, and $\mathcal{S}_{C}$ denotes the $C$-dimensional simplex.

We follow \cite{sensoy2018evidential} and construct an evidence loss $\mathcal{L}_{evi}$ to align the multi-class malware probability $p_{m}$ with the ground-truth label $y^{mul}$. The loss $\mathcal{L}_{evi}$ is formulated as follows:
\begin{equation}
\begin{aligned}
    \mathcal{L}_{evi} &=\int\left[\sum_{k=c}^{C}-y_{m}^{c} \log \left(p_m^{c}\right)\right] D(p_m \mid f_m) d p_m 
    \\
    &=\sum_{c=1}^{C} y_m^{c}\left(\psi\left({S}^c\right)-\psi\left(f_m^{c}\right)\right),
\end{aligned}
\end{equation}
where $\psi(\cdot)$ is the \textit{digamma} function. Then the overall loss $\mathcal{L}_{\mathcal{F}_{mul}}$ for the multi-class module $\mathcal{F}_{mul}$ is formulated as:

\begin{equation}
\mathcal{L}_{\mathcal{F}_{\text{mul}}} = \mathcal{L}_{\text{CE}} + \lambda_1\mathcal{L}_{evi},
\end{equation}
where $\lambda_1$ is a balancing hyperparameter that prevents any loss component from dominating, and $\mathcal{L}_{ce}$ is the cross-entropy loss, which can be defined as:
\begin{equation}
\mathcal{L}_{\text{CE}} = -\sum_{c=0}^C y^{\text{mul}}_{c} \log(f_{\text{m}}^{c}).
\end{equation}


Next, we attempt to construct the latent relationship within the multi-class logits and use the binary label to optimize $\mathcal{F}_{\text{bin}}$. In this phase, $\mathcal{F}_{\text{mul}}$ remains frozen, while $\mathcal{F}_{\text{bin}}$ is trained using a hybrid loss, defined as follows:
\begin{equation}
\label{eq7}
\mathcal{L}_{\mathcal{F}_{\text{bin}}} = \mathcal{L}_{\text{supcon}} + \lambda_2\mathcal{L}_{\text{BCE}},
\end{equation}
where $\lambda_2$ is a balancing hyperparameter. $\mathcal{L}_{\text{BCE}}$ is the binary cross-entropy loss, formulated as:
\begin{equation}
\label{eq8}
\scalebox{0.9}{$
\mathcal{L}_{\text{BCE}} = -\sum_{i=1}^M \left[y_i^{\text{bin}}\log(\mathcal{F}_{\text{bin}}(x_i)) + (1-y_i^{\text{bin}})\log(1-\mathcal{F}_{\text{bin}}(x_i))\right].
$}
\end{equation}
Additionally, $\mathcal{L}_{\text{supcon}}$ represents the supervised contrastive loss:
\begin{equation}
\label{eq9}
\mathcal{L}_{\text{supcon}} = \sum_{i=1}^M \frac{-1}{|P(i)|}\sum_{k\in P(i)}\log\frac{\exp(z_i\cdot z_k/\tau)}{\sum_{a\in A(i)}\exp(z_i\cdot z_a/\tau)},
\end{equation}
where $z_i = \mathcal{F}_{\text{bin}}(x_i)$ denotes the normalized feature embedding of sample $x_i$ in the feature space of the binary classifier $\mathcal{F}_{\text{bin}}$. The temperature hyperparameter $\tau$ controls the sharpness of the similarity distribution. The set $P(i)$ contains positive samples (including $x_i$ itself) that share the same binary label $y^\text{bin}_i$ as $x_i$, while the set $A(i)$ contains negative samples with different binary labels. The loss function encourages the model to increase the similarity between positive pairs while reducing it between negative pairs.


\noindent \textbf{Detector $\mathcal{F}_{\text{D}}$:} 
Simultaneously, the detector $\mathcal{F}_{\text{D}}$ consists of an encoder $\mathcal{F}_{\text{enc}}:\mathcal{X}\to\mathbb{R}^d$ and a classifier $\mathcal{F}_{\text{cls}}:\mathbb{R}^d\to\mathbb{R}^2$, which are jointly trained as an integrated module. 
To address the inherent class imbalance in Android malware detection datasets and enhance the classifier's discriminative power, the detector employs a weighted composite loss function:
\begin{equation}
\label{eq10}
\mathcal{L}_{\mathcal{F}_{\text{D}}} = \mathcal{L}_{\text{supcon}} + \lambda_3\mathcal{L}_{\text{w-bce}},
\end{equation}
$\lambda_3$ also is a balancing hyperparameter of $\mathcal{L}_{\mathcal{F}_{\text{D}}}$, where $\mathcal{L}_{\text{w-bce}}$ represents the weighted binary cross-entropy loss:
\begin{equation}
\mathcal{L}_{\text{w-bce}} = \sum_{i=1}^M w_i\left[y^{\text{bin}}_i\log(p_i^D) + (1-y_i^{\text{bin}})\log(1-p_i^D)\right],
\end{equation}
where $p_i^D =\mathcal{F}_{\text{D}}(x_i)= \mathcal{F}_{\text{cls}}(\mathcal{F}_{\text{enc}}(x_i))$ denotes the classifier's prediction, and $w_i$ represents the sample-specific weight. 

The same supervised contrastive loss $\mathcal{L}_{\text{supcon}}$ (Equation \eqref{eq9} ) as used for the sampler is employed, where $z_i = \mathcal{F}_{\text{enc}}(x_i)$ denotes the normalized feature embedding, with $\tau$, $P(i)$, and $A(i)$ defined analogously to the sampler's contrastive loss. This formulation maintains robustness against class imbalance while learning discriminative representations resilient to evolving threats and distribution shifts.

\noindent \textbf{Codebook $\mathcal{V}$:}
After training the detector, we initialize our vector codebook $\mathcal{V}$ by extracting 512-dimensional feature vectors from the penultimate layer of the encoder $\mathcal{F}_{\text{enc}}$, sorted according to classifier confidence scores. The vector codebook stores tuples consisting of a label, vector, and confidence score. Specifically, the codebook contains a total of 50 benign samples ($N_{\text{benign}} = 50$) and up to 3 samples per malware family ($N_{\text{mal}} = 3$), where $N_{\text{benign}}$ denotes the number of benign samples and $N_{\text{mal}}$ represents the maximum number of samples per malware family.

Inspired by the concept of prototypes and the ETF framework in prototype learning, our goal is to enhance the differentiation between benign and malware by increasing the inter-class distance while simultaneously minimizing the intra-class distance within each class. To achieve this, we perform two key operations on the feature vectors stored in the codebook. 

First, we compute the centroids of the feature vectors for each class (benign and malware) as follows:
 
\begin{equation}
\mathbf{v}^{\text{centroid}}_j = \frac{1}{N_j} \sum_{i=1}^{N_j} \mathbf{v}_j^i,
\end{equation}
where $\mathbf{v}_j^i$ is the feature vector of the $i$-th sample in class $j$, and $N_j$ is the number of samples in class $j$, which can belong to either $N_{\text{benign}}$ or up to $N_{\text{mal}}$.

After calculating the feature vector centroids, we update the feature vectors to bring them closer to their respective centroids:

\begin{equation}
{\mathbf{v}_j^i}' = \mathbf{v}_j^i + (\theta_1 + \theta_2 \cdot s_\text{confidence}) \cdot (\mathbf{v}^{\text{centroid}}_j - \mathbf{v}_j^i),
\end{equation}
where $\mathbf{v}_j^i$ is the current feature vector, $\theta_1$ and $\theta_2$ are hyperparameters that control the pull strength toward the $j$-th class's centroid $\mathbf{v}^{\text{centroid}}_j$, and $s_\text{confidence}$ represents the confidence score.
This operation ensures that the feature vectors of the benign class $\mathbf{v}_{\text{benign}}^0$ and each malware family class $\mathbf{v}_{\text{mal}}^i$ become more compact within their respective classes, moving closer to their centroids.

Next, to enhance the separation between the benign and malicious classes, we try to orthogonalize the feature vector of the benign and malicious classes. This is achieved by updating the feature vectors of the malware samples as follows:

\begin{equation}
{\mathbf{v}^i_j}' = \mathbf{v}^i_j - \theta_3 \cdot cos(\mathbf{v}^i_j ,\mathbf{v}^{\text{centroid}}_0) \cdot \mathbf{v}^{\text{centroid}}_0,
\end{equation}
where $\mathbf{v}^i_j$ is the feature vector of the malware sample, $\mathbf{v}^{\text{centroid}}_0$ is the centroid of the benign class, and $\theta_3$ is a hyperparameter that controls the separation intensity. 
This operation makes the malware feature vectors more orthogonal to the benign class vectors, thereby increasing inter-class separation and enhancing the model's ability to distinguish between benign and malicious samples.

\begin{figure*}[t]
    \centering
    \includegraphics[width=\textwidth]{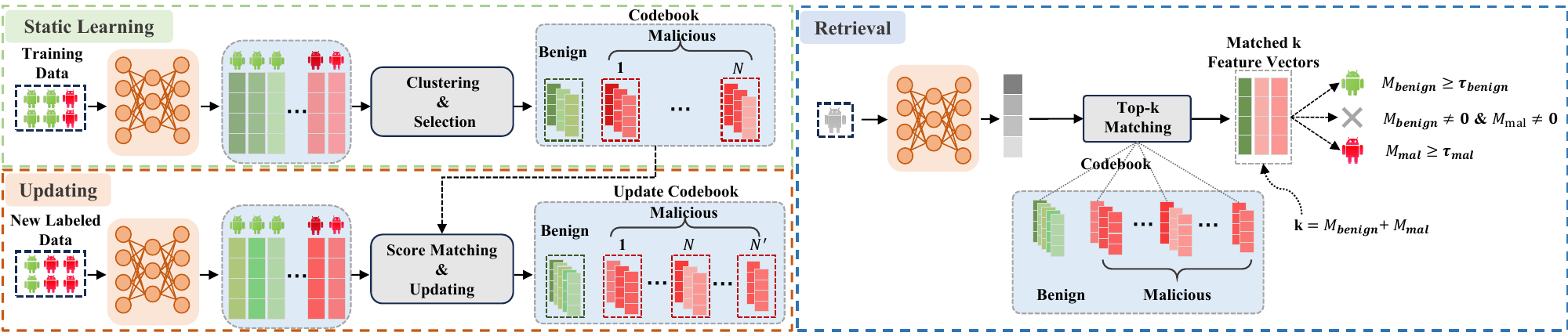}
    \caption{The pipeline of the retrieval framework.}
    \label{fig2}
\end{figure*}


\subsection{Continual Learning}
The continual learning stage consists of two key phases: sampling and learning. This approach is designed to address the challenges of catastrophic forgetting and sample imbalance in Android malware detection tasks involving time-continuous sequences. The ultimate goal is to maintain the model's performance for malware detection.

\noindent \textbf{Sampling Phase:} 
For each incoming monthly test batch $\mathcal{D}_n$, $\mathcal{F}_{\text{S}}$ calculates uncertainty scores through its cascaded modules: 
\begin{enumerate}
    \item The multi-class module $\mathcal{F}_{\text{mul}}$ computes malware family-aware uncertainty scores $\mathbf{s}_{\text{mul}} = \mathcal{F}_{\text{mul}}(\mathcal{D}_n)$ to preserve long-term malware family knowledge.
    \item The adaptive binary module $\mathcal{F}_{\text{bin}}$ generates boundary-sensitive scores $\mathbf{s}_{\text{bin}} = \mathcal{F}_{\text{bin}}(\mathcal{D}_n)$ via contrastive learning between benign and malware samples.
\end{enumerate}
The system selects the top-$k$ samples using a hybrid strategy:
\begin{equation}
    \mathcal{D}_n' = \underset{x \in \mathcal{D}_n}{\text{top}_{\mu B}} ( \mathbf{s}_{\text{mul}}(x)) + \underset{x \in \mathcal{D}_n}{\text{top}_{(1-\mu) B}} ( \mathbf{s}_{\text{bin}}(x)),
\end{equation}
where $B$ is the fixed budget and $\mu = 0.5$ balances the contributions of each module. A minimum of 10\% benign quota is enforced by replacing low-scoring malware samples when benign instances are underrepresented. The selected samples, $\mathcal{D}_n'$, are labeled by human experts and added to the memory bank $\mathcal{M}$.

\noindent \textbf{Learning Phase:} 
In the learning phase, both the multi-class module in the sampler, $\mathcal{F}_{\text{S}}$, and the detector, $\mathcal{F}_{\text{D}}$, are fine-tuned. This process combines the currently selected samples, $\mathcal{D}_n'$, with part of the historical data, $\mathcal{M}$, from the memory bank (usually 20\%), fine-tuning the model through a composite objective, similar to the static learning phase. 

In this fine-tuning phase, the objective function for the sampler combines the classification loss, $\mathcal{L}_{\text{BCE}}$ (Equation \eqref{eq8}), computed on the current samples, $ \mathcal{D}_n'$, with the contrastive loss, $\mathcal{L}_{\text{supcon}}$ (Equation \eqref{eq9}), applied to the feature space of $\mathcal{F}_{\text{bin}}$. Similarly, the detector is fine-tuned using the loss function, $\mathcal{L}_{\mathcal{F}_{\text{D}}}$ (Equation \eqref{eq10}), where the data consists of the new samples dataset, $\mathcal{D}_n'$. This regularization encourages the model to maintain clear decision boundaries by enhancing the separation between positive and negative pairs, thereby refining the feature space learned in previous iterations.

While the optimization process and loss functions are largely consistent with those used in the static learning phase, notable differences exist. Specifically, the number of training epochs is significantly reduced compared to the static learning phase.

This continual learning strategy allows the model to adapt to emerging malware and benign software, ensuring sustained high performance despite shifts in data distribution. By utilizing a few current and historical samples, the model maintains its performance while preserving knowledge from previous iterations.

\subsection{Retrieval-based Malware Detection}
In Android malware detection, traditional deep learning approaches face two key limitations: decision boundaries that favor majority-class samples, leading to missed detections of novel variants, and architectures that cannot adapt to evolving malware characteristics. To overcome these issues, we introduce a feature vector retrieval and matching module that enhances sensitivity to minority samples while improving generalization to unseen malware through dynamic similarity matching.
The pipeline of the proposed retrieval-based malware detection framework is illustrated in Fig.~\ref{fig2}.

\subsubsection{Construction and Updating of Codebook}
The details of the codebook construction have been mentioned earlier. To improve retrieval and matching performance, we aim to make the feature representations of each class more compact while increasing the distance between benign and malicious classes. Following codebook construction, we further refine the feature space by applying additional operations, which are described in detail in the methodology section. In this space, the angular relationships between different categories are optimized to maximize inter-class separation while ensuring intra-class compactness.

During continual learning, our codebook updating mechanism maintains an evolving but stable feature representation through three key operations. First, for newly encountered samples, we evaluate their classifier confidence scores and selectively update the codebook by replacing existing lower-confidence vectors of the same class, strictly adhering to predefined capacity constraints ($N_{\text{benign}}=50$ and $N_{\text{mal}}=3$).

We maintain the optimized geometric structure by recalculating the category centroid through mean aggregation whenever vectors are added or replaced. This ensures that the updated samples in the codebook retain strong benign-malware discrimination while preserving the relationships within each malware family.

\begin{algorithm}[t]
\caption{The learning procedure of UGSR}
\label{alg:ugsr}
\begin{algorithmic}[1]

\STATE \textbf{Static Learning:}
\STATE Initialize hierarchical sampler $\mathcal{F}_S$:
\STATE \quad step 1: Train multi-class module $\mathcal{F}_{mul}$
\STATE \quad step 2: Train binary module $\mathcal{F}_{bin}$ while frozen $\mathcal{F}_{mul}$
\STATE Initialize detector $\mathcal{F}_D$ 
\STATE Construct vector codebook $\mathcal{V}$
\WHILE{$T_n \neq  \emptyset$}
\STATE \textbf{Testing:}
\FOR{each testing period $t \in \{T_{1},T_{2}...T_{n}...\}$}
    \STATE Load $\mathcal{D}_n={\{x\}}$  
    \STATE $\mathbf{F}_n \gets \mathcal{F}_{enc}(\mathcal{D}_n)$ (\text{Feature Extraction})
    \STATE \textbf{Hybrid Prediction:}
    \FOR {each $x_i \in \mathcal{D}_n$}
        \STATE Neural network prediction ($\mathcal{F}_D$): \\ $p_i \gets \mathcal{F}_{D}(x_i)$
        \STATE Codebook similarity matching ($\mathcal{V}$): \\ $m_i \gets \text{Top-k}(\mathcal{V}, \mathbf{F}_n, k=3)$
        \STATE $\text{Fuse}(p_i, m_i)$
    \ENDFOR
\ENDFOR
\STATE \textbf{Continual Learning:}
\FOR {each evaluation period $T_n$}
    \STATE Load test samples $\mathcal{D}_{n}$
    \STATE \textbf{Active Sampling:(B is sample budget)}
    \STATE Select informative samples using $\mathcal{F}_S$:
    \STATE \quad - $\mu B$  highest uncertainty score samples($\mathcal{F}_{mul}$) 
    \STATE \quad - $(1-\mu) B$  highest uncertain score samples($\mathcal{F}_{bin}$)
    \STATE Update accumulated data $\mathcal{D}^{\prime}_n$ to memory bank $\mathcal{M}$
    \STATE \textbf{Model Update:}
    \STATE Fine-tune $\mathcal{F}_S$ and $\mathcal{F}_D$ on $\mathcal{M}$
    \STATE Update $\mathcal{V}$ with new features
    \STATE \textbf{Space Optimization:}
    \STATE Apply ETF principle to $\mathcal{V}$
\ENDFOR
\STATE \textbf{Return} $\mathcal{F}_S$, $\mathcal{F}_D$, $\mathcal{V}$
\ENDWHILE

\end{algorithmic}
\end{algorithm}

\subsubsection{Vector Retrieval and Match}

In the testing stage, the feature vector library is used to retrieve the similarity of input samples. The specific steps are as follows: the test sample $x$ is passed through the encoder $E$ to obtain a 512-dimensional vector $v_x$, after which all the feature vectors in the codebook are traversed to calculate the cosine similarity between the input feature vector $v_x$ and each feature vector $v_i$ in the codebook, $\mathcal{V} = \{v_1,v_2,\dots,v_n \}$. The cosine similarity metric is computed as follows:

\begin{equation}
\label{16}
s(v_x, v_i) = \frac{v_x \cdot v_i}{\|v_x\| \|v_i\|},
\end{equation}
where $s(\cdot,\cdot)$ denotes the similarity score ranging from -1 to 1. Retrieve the $k$ most similar feature vectors to the input feature vector based on the similarity score $s(v_x, v_i)$, where $k$ is a tunable hyperparameter. 

Based on the matching results and the classifier's original predictions, we propose a dynamic decision strategy: (1) when benign matches $M_{benign}$ exceed the threshold $\theta$ (set $\theta$ = $k$), we predict its class as benign; (2) when malicious matches $M_{mal}$ exceed $\theta$, we predict its class as malicious; (3) otherwise, we retain the classifier's prediction, $p_i^D$. We propose a dynamic decision strategy formalized as:
\begin{equation}
\label{eq17}
\hat{p}_{final} = 
\begin{cases}
    0, & \text{if } M_{\text{benign}} = \theta \\
    1, & \text{if } M_{\text{mal}} = \theta \\
    p_i^D, & \text{otherwise}
\end{cases}.
\end{equation}

This dual-threshold mechanism provides a robust fallback to the original classifier when retrieval evidence is inconclusive, while leveraging codebook matches when they provide strong discriminative signals. The top-$k$ thresholding naturally adapts to varying confidence levels across different malware families

This retrieval-enhanced decision framework combines the strengths of both parametric classifier predictions and non-parametric similarity matching, achieving more robust and interpretable Android malware detection through their synergistic integration. The approach effectively balances the stability of learned classifier patterns with the adaptability of instance-based retrieval, which is particularly valuable in handling ambiguous cases where the classifier's confidence is low.

The overall procedure is formalized in Algorithm~\ref{alg:ugsr}, where the UGSR framework sequentially executes static model initialization, hybrid prediction (combining $\mathcal{F}_D$ and codebook $\mathcal{V}$), and continual learning with uncertainty-driven sampling and ETF-based space optimization.

\section{Experiments}
\subsection{Dataset}
\begin{figure}[!t]
    \centering
    \subfloat[]{\includegraphics[width=0.64\columnwidth]{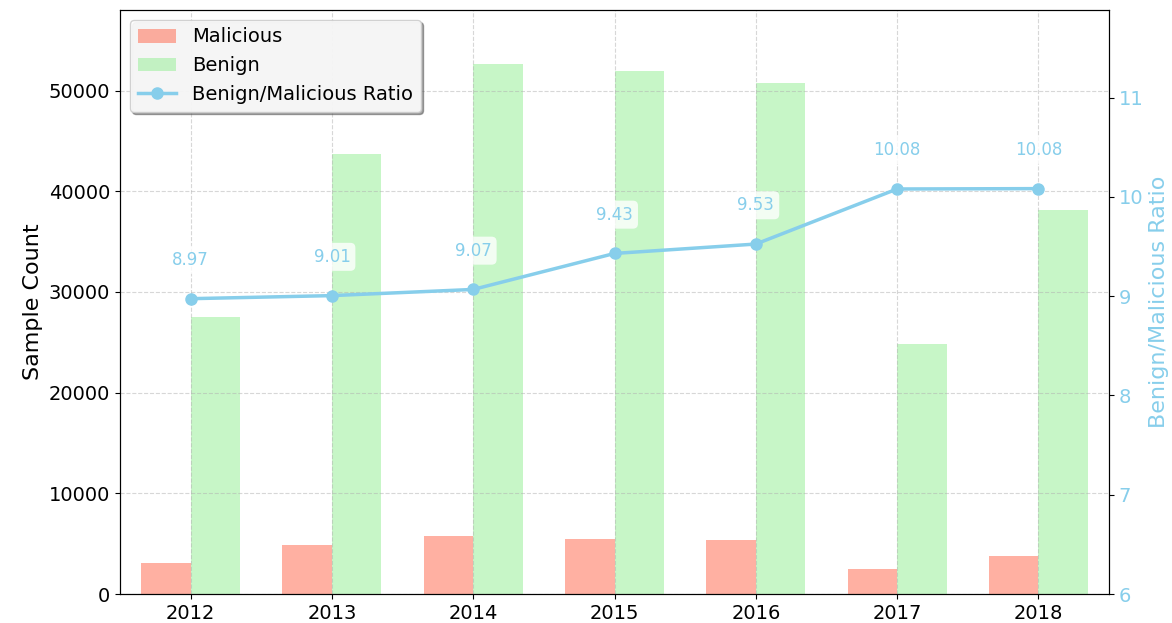}}
    \subfloat[]{\includegraphics[width=0.33\columnwidth]{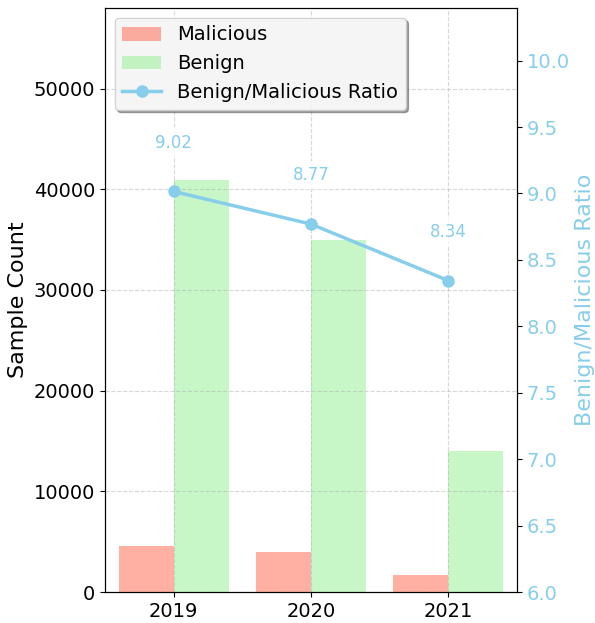}}
        
    \caption{The introduction of datasets: (a) represents the APIGraph dataset and (b) represents the Androzoo datasets.}
    \vspace{-15pt}
    \label{fig3}
\end{figure}

We conducted evaluations on two benchmark datasets, APIGraph~\cite{zhang2020enhancing} and AndroZoo~\cite{androzoo}. 
APIGraph contains about 320,000 Android applications published over seven years (2012-2018. AndroZoo contains about 100,000 Android applications published between 2019 and 2021. The ratio of benign samples to malicious samples in these datasets is approximately 9:1. 
The samples in both datasets are highly imbalanced. To make this more intuitive, the distributions of these datasets are illustrated in Fig.~\ref{fig3}.



\subsection{Experimental Settings}
In both datasets, the first 12 months of data were used as the static training set, $\mathcal{D}_s$, with the subsequent months reserved for continual learning.
The $\mathcal{F}_{\text{mul}}$ module in the sampler and $\mathcal{F}_{\text{enc}}$ module in the detector share the same MLP architecture, which consists of five linear layers. Similarly, the $\mathcal{F}_{\text{bin}}$ module in the sampler and $\mathcal{F}_{\text{cls}}$ module in the detector also share the same MLP architecture, which contains four linear layers. The only distinction between them is that the final layer of $\mathcal{F}_{\text{mul}}$ in the sampler includes a softmax activation function.

Following the work by Chen \textit{et al.}~\cite{chen2023continuous}, in the static learning stage, we trained our sampler and detector for 200 epochs using SGD (learning rate $\eta=3\times10^{-4}$) with a batch size of 1024. In the continual learning stage, we freeze $\mathcal{F}_{\text{bin}}$ and fine-tune the remaining components for 50 epochs using Adam ($\eta=5\times10^{-5}$), while maintaining a 40\% benign sample ratio in incremental batches. For the codebook, we set $N_{\text{benign}}=50$ and $N_{\text{mal}}=3$ per family, with matching thresholds set to $\tau=3$.

\subsection{Metrics}
We use the fundamental metrics of true positive rate (TPR) and true negative rate (TNR) to evaluate performance. Given the class imbalance in Android malware detection (with benign-malicious ratios approaching 9:1), standard accuracy becomes unreliable because it favors the majority class and may mask failures in detecting malware. Therefore, we adopt the balanced accuracy metric (mACC), which equally weights both classes to comprehensively evaluate performance. mACC is formulated as:
\begin{equation}
\text{mACC} = (\text{TPR} + \text{TNR}) / 2.
\end{equation}

Besides, F2 is used to evaluate the performance as follows:

\begin{equation}
F2 =  \frac{(1 + 2^2) \cdot \text{Precision} \cdot \text{Recall}}{2^2 \cdot \text{Precision} + \text{Recall}}.
\end{equation}


Given that Android malware detection is an unbalanced task, we evaluated the performance of different methods using F2. F2 is a suitable metric for Android malware detection as it emphasizes recall, which makes it more appropriate for security scenarios where identifying malware is of greater importance.

The geometric mean, G-Mean, is also used to evaluate performance, as it considers both TPR and TNR. G-Mean can be calculated as follows:
\begin{equation}
\text{G-mean} = \sqrt{\text{TPR} \times \text{TNR}}.
\end{equation}


\subsection{Baselines}
The first baseline applies active learning with base binary classifiers, including MLP, linear SVM, and Gradient Boosted Decision Trees (GBDT) \cite{zhang2020enhancing,yang2021bodmas}. The second baseline utilizes an SVM classifier guided by CADE OOD scores \cite{yang2021cade}, which were originally designed for distribution drift detection and also support the sample labeling strategy employed in our experiments. Lastly, the third baseline incorporates pseudo-loss scores for sample selection \cite{chen2023continuous}.

\begin{table}[!t]
\centering
\caption{Experimental Results on APIGraph Dataset.}
\label{tab1}
\resizebox{\columnwidth}{!}{
\begin{tabular}{llccccc}
\toprule
\multirow{2}{*}{Method}                                 & \multicolumn{5}{c}{Average Performance (\%)} \\ \cmidrule(l){2-6} 
 & TPR    & TNR     & F2     & G-mean  & mACC  \\ \midrule
Binary MLP       & 76.23  & \textbf{99.48}     & 79.26  & 87.08 & 87.86  \\
Multiclass MLP   & 83.90  & 95.36      & 79.81  & 89.45  & 89.63 \\
Binary SVM       & 83.08  & 99.39    & 85.02  & 90.87  & 91.24 \\
CADE OOD\cite{yang2021cade} & 63.89  & 87.10    & 55.08  & 74.60  & 75.50  \\
Binary GBDT\cite{yang2021bodmas}& 68.25  & 99.46  & 72.13  & 82.39  & 83.86 \\
USENIX23\cite{chen2023continuous} & 84.85  & \textbf{99.48}   & 86.66  & 91.87   & 92.17 \\
UGSR & \textbf{92.95}  & 95.58  & \textbf{86.82}  & \textbf{94.22}  & \textbf{94.26}  \\ \bottomrule
\end{tabular}
}
\end{table}

\begin{table}[!t]
\centering
\caption{Experimental Results on AndroZoo Dataset.}\label{tab2}
\resizebox{\columnwidth}{!}{
\begin{tabular}{llcccccc}
\toprule
\multirow{2}{*}{Method}                                   & \multicolumn{5}{c}{Average   Performance (\%)} \\ \cmidrule(l){2-6} 
   & TPR     & TNR    & F2     & G-mean   & mACC   \\ \midrule
Binary   MLP     & 46.88   & 99.54   & 51.97  & 68.31   & 73.21  \\
Multiclass   MLP  & 50.14   & 71.48   & 35.47  & 53.56   & 60.81  \\
Binary   SVM       & 51.23   & \textbf{99.71}    & 56.44  & 71.47  & 75.47   \\
CADE OOD\cite{yang2021cade} & 37.99   & 99.45   & 42.88  & 61.47  & 68.72  \\
Binary   GBDT\cite{yang2021bodmas}    & 49.65   & 99.53   & 54.69  & 70.29   & 74.59  \\
Usenix23\cite{chen2023continuous}   & 72.35   & 99.47     & 75.82  & 84.83  & 85.91  \\
UGSR    & \textbf{80.56}   & 96.01   & \textbf{77.32}  & \textbf{86.68}  & \textbf{88.29} \\ \bottomrule
\end{tabular}
}
\label{tab:androzoo}
\end{table}

\subsection{Results}
We apply a monthly budget of 50 samples for analyst labeling to ensure a fair comparison across all methods. The results in APIGraph and AndroZoo can be found in Tables~\ref{tab1} and~\ref{tab2}, respectively. It can be seen that the other methods tend to overfit to the majority benign class, leading to low TPR and high TNR. In contrast, the results demonstrate the strong capability of UGSR in detecting the minority class. In terms of overall performance metrics, our method outperforms all other baselines. The reason lies in that our method prioritizes the most uncertain minority class samples through uncertainty-driven hierarchical sampling, effectively preventing overfitting to the majority class.

UGSR's TPR and mACC consistently outperform all baselines across both the APIGraph and AndroZoo datasets. It demonstrates a strong capability in detecting rare malicious samples while maintaining balanced classification. Notably, UGSR also achieves a higher F2 score in both cases, consistent with the fact that failing to detect malware is significantly more costly than false alarms.

Compared to baseline methods that tend to overfit to the majority benign class, UGSR improves the detection of minority malicious samples with only a slight reduction in benign classification accuracy. This balance leads to superior G-mean and mACC scores, reflecting the model’s robustness in handling class imbalance.

Furthermore, UGSR achieves this performance with a substantially lower annotation cost. While previous optimal work~\cite{chen2023continuous} required 200 labeled samples to reach a 90\% TPR on the APIGraph dataset, UGSR achieves comparable or better performance using only 50 labeled samples per fine-tuning iteration. This highlights its efficiency and practicality for real-world, continuously evolving Android malware detection systems.

\begin{table}[]
\centering
\small
\scriptsize
\caption{Few-shot Continual Learning results on the APIGraph Dataset.}
\label{tab3}
\begin{tabular}{llcccccc}
\toprule
\multirow{2}{*}{Model}  & \multirow{2}{*}{\shortstack{Sample \\ Budget}}& \multicolumn{5}{c}{Average Performance (\%)}                                       \\ \cmidrule(l){3-7} 
&        & TPR       & TNR     & F2      & G-mean   & mACC  \\ \midrule
USENIX23\cite{chen2023continuous}  & Count = 10   & 77.46          & \textbf{99.25}         & 79.99          & 87.68       & 88.36     \\
Ours  & Count = 10   & \textbf{89.45} & 95.46     & \textbf{83.92} & \textbf{92.33}   & \textbf{92.46}\\ \midrule 
USENIX23\cite{chen2023continuous}                                                     & Count = 2    & 67.63    & \textbf{99.13}    & 71.12          & 81.88        & 83.38   \\
Ours   &  Count = 2    & \textbf{81.62} & 96.14 & \textbf{79.03} & \textbf{88.58}  & \textbf{88.88}   \\ \bottomrule
\end{tabular}
\end{table}

\begin{table}[!t]
\centering
\small
\scriptsize
\caption{Few-shot Continual Learning results on the AndroZoo Dataset.}
\label{tab4}
\begin{tabular}{llcccccc}
\toprule
\multirow{2}{*}{Model} & \multirow{2}{*}{\shortstack{Sample \\ Budget}} & \multicolumn{5}{c}{Average Performance (\%)}                                       \\ \cmidrule(l){3-7} 
  &   & TPR    & TNR     & F2     & G-mean    & mACC       \\ \midrule
USENIX23\cite{chen2023continuous}                                                   & Count = 10    & 67.07   & \textbf{99.48}  & 71.09          & 81.68    & 83.27    \\
Ours      & Count = 10          & \textbf{75.41} & 96.24        & \textbf{73.47} & \textbf{84.14}  & \textbf{85.83} \\  \midrule 
USENIX23\cite{chen2023continuous}                                    & Count = 2   & 59.54    & \textbf{99.12}     & 62.44          & 74.75     & 79.33     \\
Ours      & Count = 2          & \textbf{62.67} & 97.17     & \textbf{62.74} & \textbf{75.31} & \textbf{79.92} \\ \bottomrule
\end{tabular}
\end{table}

Additionally, we conducted experiments on the APIGraph dataset under a few-shot setting, with the sample budget set to 10 and 2, respectively. In this experiment, we treat the USENIX23 as the baseline, and the results can be found in Table~\ref{tab3}. Our method achieves a more than 10\% increase in TPR with only a minor reduction in TNR. Moreover, improvements are observed across nearly all key metrics, which demonstrate that our method can effectively maintain balanced and reliable detection even under limited sample budgets. Similar results are observed in AndroZoo, as shown in Table~\ref{tab4}. These results further demonstrate the effectiveness of our approach, highlighting its robustness under the small sample budget setting.

\begin{figure}[!t]
\centering
\includegraphics[width=\columnwidth]{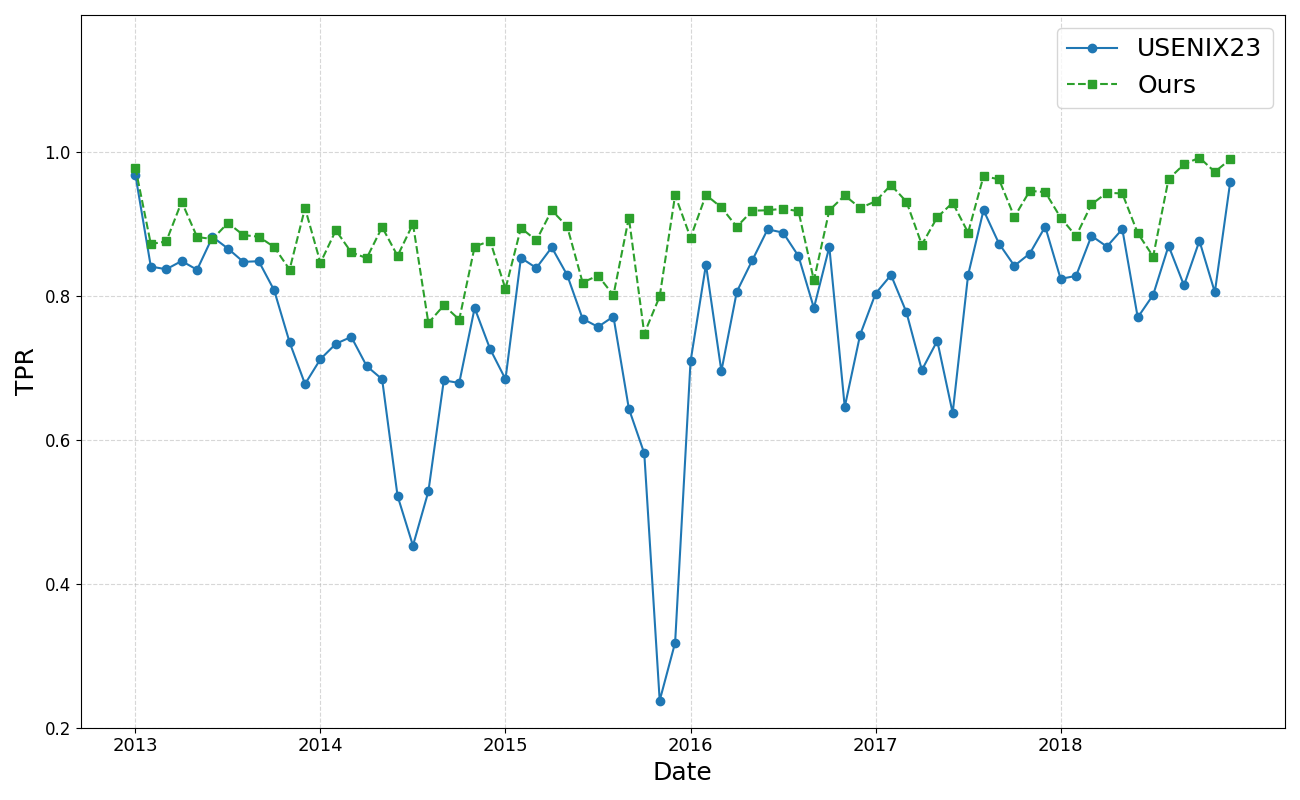}
\caption{Performance comparison between our method and USENIX23\cite{chen2023continuous}.} 
\label{fig4}
\end{figure}

To intuitively compare performance across different months, Fig.~\ref{fig4} shows the TPR changes on the APIGraph dataset when the sample budget is set to 10. It can be seen that our method is more robust than the baseline.
Our method exhibits minimal fluctuation across different months, highlighting its robustness. In contrast, the baseline is highly sensitive to the data, with TPR reaching only about 20\% in some months. The main reason lies in that our method not only selects the most uncertain samples for continual learning but also utilizes a codebook that retains representative samples from previously encountered benign and malware families, preserving their features.

\subsection{Ablation Experiments}
\begin{figure}
\includegraphics[width=\columnwidth]{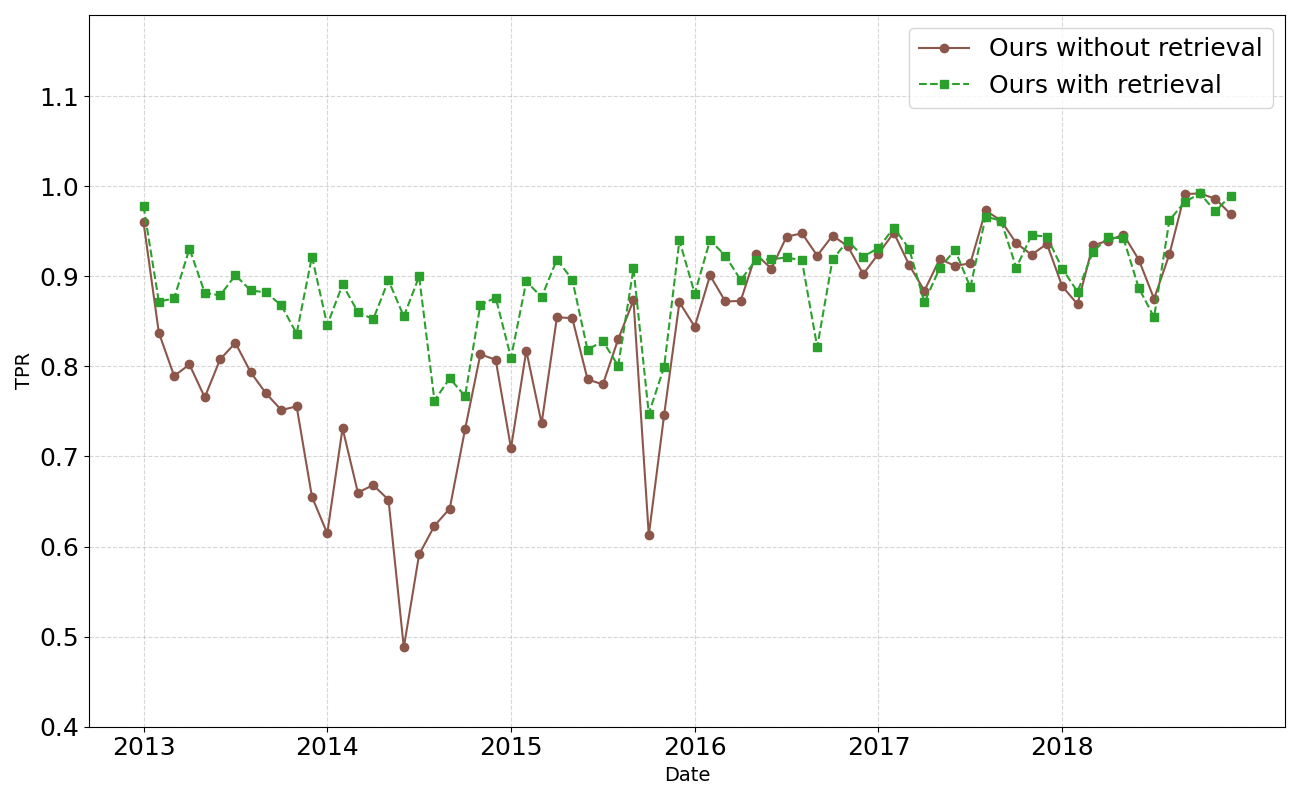}
\caption{Ablation study about the retrieval module. The horizontal and vertical axes represent the time series of the dataset and the TPR of malware detection, respectively.} 
\label{fig5}
\end{figure}

In this subsection, to evaluate the effectiveness of transforming the traditional classification-based detection paradigm into a retrieval-based detection paradigm, we conducted related ablation experiments. The monthly results can be found in Fig.~\ref{fig5}, where it can be seen that the retrieval module effectively stabilizes performance and prevents oscillation. For the average performance shown in Table~\ref{tab5}, TPR increases from 83.75\% to 89.45\%, accompanied by substantial gains in mACC and G-mean. 


\begin{table}[t]
\centering
\caption{Effect of the proposed vector retrieval module.}
\label{tab5}
\begin{tabular}{@{}l@{\hspace{0.1em}}c@{\hspace{0.5em}}cccccc@{}}
\toprule
\multirow{2}{*}{Model} & \multirow{2}{*}{\begin{tabular}[c]{@{}l@{}}Samples \\ Budget\end{tabular}} & \multicolumn{5}{c}{Average Performance (\%)} & \\ \cmidrule(l){3-7} 
& & TPR & TNR  & F2 & G-mean & mACC\\ \midrule
w/o retrieval & 10 & 83.75 & 91.14 & 74.31 & 87.37& 87.45 \\
w retrieval   & 10  & \textbf{89.45} & \textbf{95.46}  & \textbf{83.92} & \textbf{92.33}  & \textbf{92.46} \\  \bottomrule
\end{tabular}
\end{table}

\begin{table}[!t]
\caption{Effect of selector $\mathcal{F}_{mul}$.}
\label{tab6}
\begin{tabular}{@{}l@{\hspace{0.5em}}c@{\hspace{0.5em}}cccccc@{}}
\toprule
\multirow{2}{*}{Model} & \multirow{2}{*}{\begin{tabular}[c]{@{}l@{}}Samples \\ Budget\end{tabular}} & \multicolumn{5}{c}{Average Performance (\%)}                                       &                \\ \cmidrule(l){3-7} 
 &       & TPR     & TNR     & F2     & G-mean   & mACC     \\ \midrule
w/o $\mathcal{F}_{mul}$    &  10        & 73.43          & 89.06                      & 56.73          & 80.87      & 87.45      \\
w $\mathcal{F}_{mul}$                          &  10                      & \textbf{89.45} & \textbf{95.46}  & \textbf{83.92} & \textbf{92.33} & \textbf{92.46}\\ \bottomrule
\end{tabular}
\end{table}

\begin{figure}[!t]
\centering
\includegraphics[width=\columnwidth]{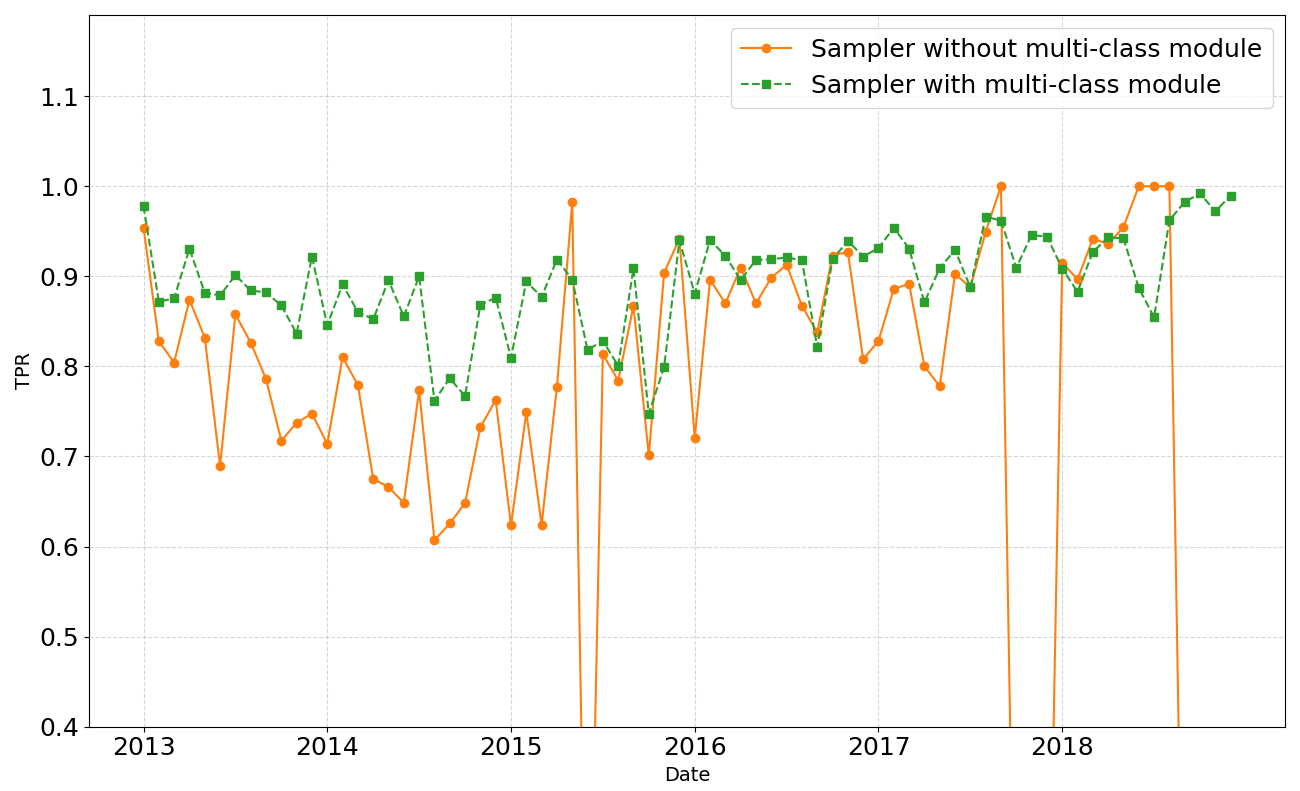}
\caption{Ablation study of sampler's multi-class module $\mathcal{F}_{mul}$. The horizontal and vertical axes represent the time series of the dataset and the TPR of malware detection, respectively.} 
\label{fig6}
\end{figure}

We further evaluated the effectiveness of the hierarchical sampler, $\mathcal{F}_{\text{mul}}$. As evidenced by Fig.~\ref{fig6} and Table~\ref{tab6}, the ablation of $\mathcal{F}_{\text{mul}}$ induces severe continual learning instability, manifested through: (1) erratic TPR oscillations with peak-to-valley variations exceeding 16.02\%, and (2) recurrent collapse events where TPR plummets to zero—a clear indicator of catastrophic overfitting wherein the model degenerates into predicting all instances as benign.

In contrast, the full model equipped with $\mathcal{F}_{\text{mul}}$ maintains consistently high TPR throughout the 2013–2018 period, even under significant distribution shifts. These findings underscore the importance of jointly leveraging both the binary selector, $\mathcal{F}_{\text{bin}}$, and the multi-class selector, $\mathcal{F}_{\text{mul}}$, to ensure stable performance. Overall, the proposed approach demonstrates superior efficacy in prioritizing the latest and uncertain samples for annotation. This strategy not only optimizes labeling budget allocation but also enhances model performance while mitigating overfitting in low-sample, non-stationary scenarios.

\begin{table}[]
\caption{Effect of Budget sample counts.}
\centering
\label{tab7}
\begin{tabular}{@{}lcccccc@{}}
\toprule
\multirow{2}{*}{\begin{tabular}[c]{@{}c@{}}Samples  Budget\end{tabular}} & \multicolumn{5}{c}{Average Performance (\%)} &        \\ \cmidrule(l){2-7}              & TPR     & TNR   & F2     & G-mean  & mACC  \\ \midrule
Count = 2    & 62.67   & 97.17   & 62.74  & 75.31  & 79.92 \\
Count = 10   & 75.41   & 96.24   & 73.47  & 84.14  & 85.83  \\
Count = 25   &  77.31  & 95.47   & 73.87  & 84.32  & 86.39 \\
Count = 50   & 80.56   & 96.01   & 77.32  & 86.68  & 88.29\\
Count = 100  & 84.69   & 93.40   & 78.16  & 88.44  & 89.04  \\
\bottomrule
\end{tabular}
\label{sample budget}
\end{table}

We further investigated the impact of varying sample budgets during the continual learning phase on the AndroZoo dataset. As shown in Table~\ref{tab7}, increasing the number of sampled data points from 2 to 10 leads to significant improvements across all metrics. Furthermore, it can be seen that our method can achieve promising performance with just a budget of 10.



\section{Conclusion}

In this paper, we present a novel uncertainty-driven hierarchical sampling framework for continual Android malware detection under data imbalance. Our dual-level architecture decouples uncertainty-guided sample selection from memory-augmented classification, enabling effective adaptation to concept drift through time-series update-based retrieval. The proposed hierarchical uncertainty quantification and dynamic vector retrieval mechanisms provide robust detection of evolving polymorphic variants while maintaining efficiency in data-scarce environments. Key innovations include an adaptive temporal retrieval engine and an optimized vector codebook,  which offer a practical solution for sustainable Android malware detection. Future work will focus on continual learning strategies for zero-day malware adaptation and federated continual learning for collaborative threat detection.

\newpage

\section{Biography Section}
 


\begin{IEEEbiography}[{\includegraphics[width=1in,height=1.25in, clip,keepaspectratio]{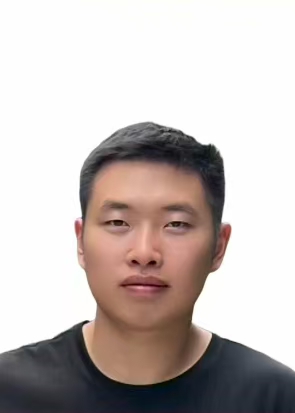}}]{Yi Xie} received the B.S. degree in Artificial Intelligence from Wuhan University of Technology, Wuhan, China, in 2023. He is currently pursuing the M.S. degree with the School of Cyber Science and Engineering at Sichuan University. His research interests include adversarial attacks on AI models and malware detection. \end{IEEEbiography}

\begin{IEEEbiography}[{\includegraphics[width=1in,height=1.25in, clip,keepaspectratio]{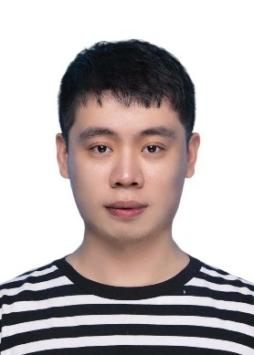}}]{Ziyuan Yang} received his Ph.D. from Sichuan University in 2025. He was a research intern at the Centre for Frontier AI Research, Agency for Science, Technology and Research (A*STAR), Singapore. In the last few years, he has published over 30 papers in leading machine learning conferences and journals, including CVPR, IJCV, TIFS, TNNLS, TCSVT, etc. He was the reviewer for leading journals or conferences, including TPAMI, TIP, TIFS, TMI, CVPR, ICCV, etc. \end{IEEEbiography}

\begin{IEEEbiography}[{\includegraphics[width=1in,height=1.25in, clip,keepaspectratio]{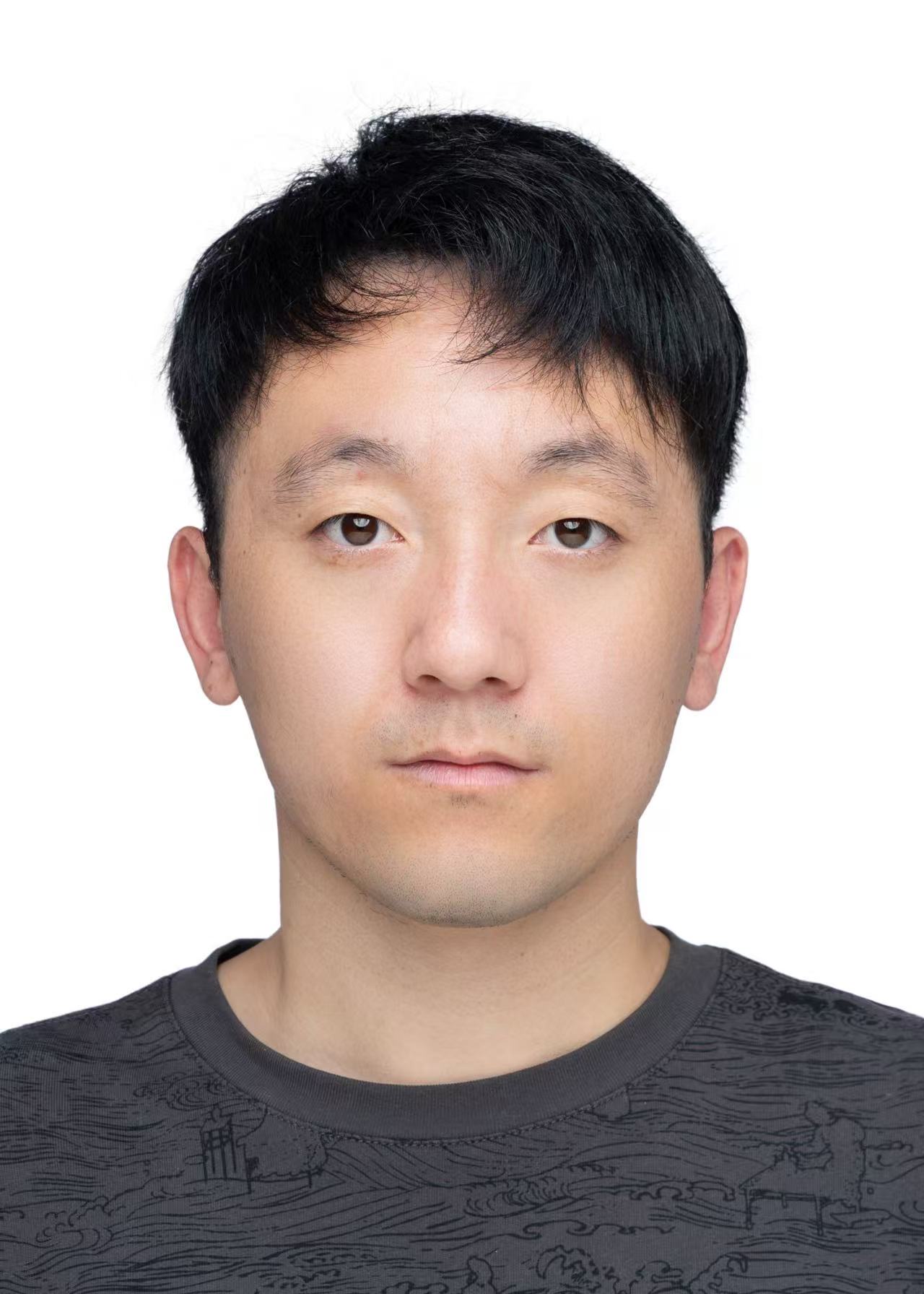}}]{Yongqiang Huang} received the M.S. degree from Sichuan University, Chengdu, China, in 2021. He is currently a PhD candidate with the School of Cyber Science and Engineering at Sichuan University. His current research interests include data privacy preservation and generative AI. \end{IEEEbiography}

\begin{IEEEbiography}[{\includegraphics[width=1in,height=1.25in, clip,keepaspectratio]{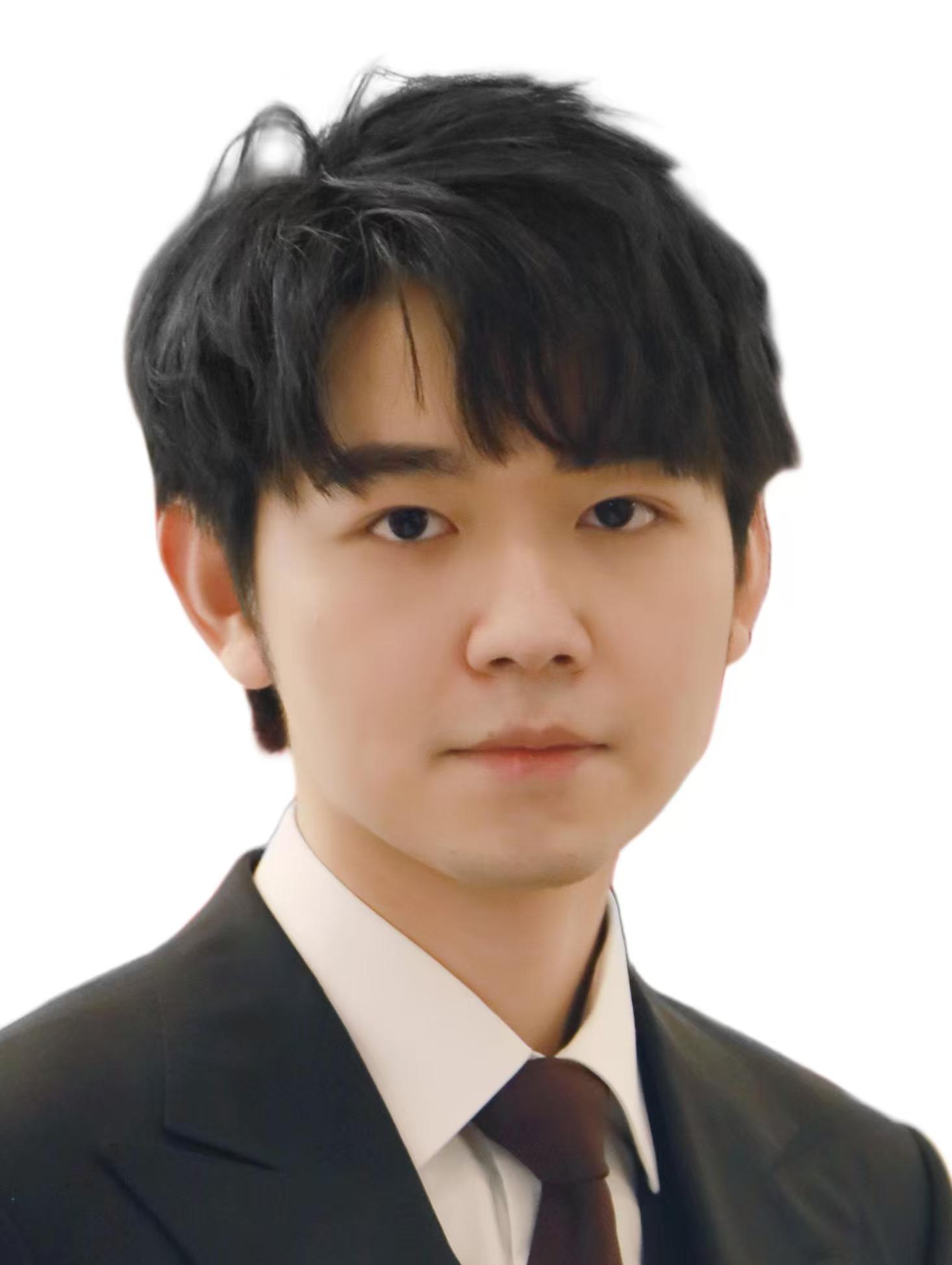}}]{Yinyu Chen} received the B.S. degree in Software Engineering from the College of Software Engineering, Sichuan University, Chengdu, China. He is currently pursuing his Ph.D. degree with the College of Computer Science, Sichuan University, Chengdu, China. His research interests include machine learning, deep learning, and medical image analysis.\end{IEEEbiography}

\begin{IEEEbiography}[{\includegraphics[width=1in,height=1.25in, clip,keepaspectratio]{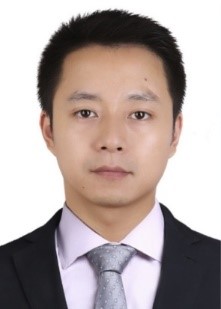}}]{Lei Zhang} received the M.S. and Ph.D. degrees in computer science and technology from Sichuan University, Chengdu, Sichuan, China. He is currently an assistant researcher in the School of Cyber Science and Engineering, Sichuan University. His research focuses on malware detection and analysis.\end{IEEEbiography}

\begin{IEEEbiography}[{\includegraphics[width=1in,height=1.25in, clip,keepaspectratio]{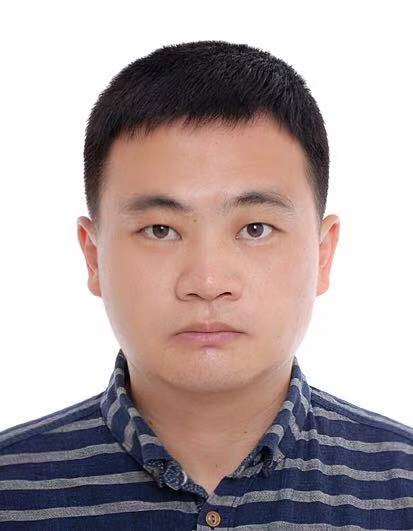}}]{Liang Liu} received the M.S. and Ph.D. degrees from Sichuan University, Chengdu, China, in 2009 and 2021, respectively. He is currently an Associate Professor with the School of Cyber Science and Engineering at Sichuan University. His current research interests include malware detection, network security, system security, and artificial intelligence. \end{IEEEbiography}

\begin{IEEEbiography}[{\includegraphics[width=1in,height=1.25in, clip,keepaspectratio]{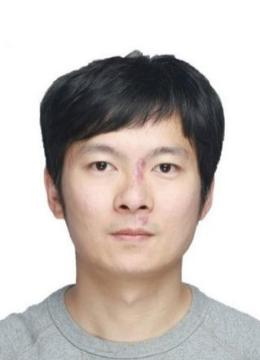}}]{Yi Zhang} (Senior Member, IEEE) received the B.S., M.S., and Ph.D. degrees in computer science and technology from the College of Computer Science, Sichuan University, Chengdu, China, in 2005, 2008, and 2012, respectively. From 2014 to 2015, he was with the Department of Biomedical Engineering, Rensselaer Polytechnic Institute, Troy, NY, USA, as a Postdoctoral Researcher. He is currently a Full Professor at the School of Cyber Science and Engineering, Sichuan University, and is the Director of the Deep Imaging Group (DIG). His research interests include medical imaging, compressive sensing, and deep learning. He authored more than 80 papers in the field of image processing. These papers were published in several leading journals, including IEEE TRANSACTIONS ON MEDICAL IMAGING, IEEE TRANSACTIONS ON COMPUTATIONAL IMAGING, Medical Image Analysis, European Radiology, Optics Express, etc., and reported by the Institute of Physics (IOP) and during the Lindau Nobel Laureate Meeting. He received major funding from the National Key R\&D Program of China, the National Natural Science Foundation of China, and the Science and Technology Support Project of Sichuan Province, China. He is a Guest Editor of the International Journal of Biomedical Imaging, Sensing and Imaging, and an Associate Editor of IEEE TRANSACTIONS ON MEDICAL IMAGING and IEEE TRANSACTIONS ON RADIATION AND PLASMA MEDICAL SCIENCES. \end{IEEEbiography}

\vfill
\end{document}